\documentclass{article}
\usepackage{spconf,graphicx}

\usepackage[english]{babel}

\usepackage{ifpdf}

\usepackage{cite} 
\usepackage{url}
\usepackage{hyperref}

\usepackage{color}
\usepackage{pgf, tikz, pgfplots}
\usetikzlibrary{shapes, arrows, automata, plotmarks}
\usetikzlibrary{calc,hobby,decorations}

\ninept

\usepackage[safe]{tipa}
\usepackage[cmex10]{amsmath}
\usepackage{amsfonts, amssymb, amsthm}
\usepackage{mathrsfs}
\usepackage{dsfont}

\usepackage{algorithm,algpseudocode}
\algnewcommand{\LeftComment}[1]{\Statex \(\triangleright\) #1}

\usepackage{enumerate}
\usepackage{multirow}
\usepackage{rotating}
\usepackage{breqn}
\usepackage{soul}
\usepackage{subcaption}
\usepackage{needspace}

\usepackage{amsmath}

\input{mySymbol.sty}
\input{pennColors.sty}

\newtheorem{lemma}{\hspace{0pt}\bf Lemma}

\newtheorem{definition}{\hspace{0pt}\bf Definition}

\title{Learning with Multigraph Convolutional Filters}

\name{Landon Butler$^{\star}$ \qquad Alejandro Parada-Mayorga$^{\dagger}$ \qquad Alejandro Ribeiro$^{\dagger}$ \thanks{All the authors contributed equally. Supported by
the National Science Foundation Graduate Research Fellowship under Grant
No. DGE-2146752.}}

\address{$^*$ University of California, Berkeley \\
$^\dagger$ University of Pennsylvania}
\begin{document}
%
\maketitle

\begin{abstract}

In this paper, we introduce a convolutional architecture to perform learning when information is supported on multigraphs. Exploiting algebraic signal processing (ASP), we propose a convolutional signal processing model on multigraphs (MSP). Then, we introduce multigraph convolutional neural networks (MGNNs) as stacked and layered structures where information is processed according to an MSP model. We also develop a procedure for tractable computation of filter coefficients in the MGNN and a low cost method to reduce the dimensionality of the information transferred between layers. We conclude by comparing the performance of MGNNs against other learning architectures on an optimal resource allocation task for multi-channel communication systems.

\end{abstract}


\begin{keywords}
Multigraphs, multigraph signal processing, multigraph convolutional 
neural networks (MGNNs), algebraic signal processing (ASP), algebraic signal model (ASM).
\end{keywords}




\section{Introduction}\label{sec:intro}


Graphs have become essential tools to capture structure within arbitrary data sets~\cite{sporns2022graph, dijkstra2019networks}. However, they are insufficient for the modeling of heterogeneous domains~\cite{meng2016interplay,9072356}. In these scenarios, \textit{multigraphs} have emerged as the natural extension of graphs~\cite{PhysRevLett.110.028701}. Additionally, convolutional information processing on graphs has been performed exploiting tools from graph signal processing (GSP); however, when applied to multigraphs, GSP allows the modeling of multiple types of information diffusion separately, but is not able to support heterogeneous mixings~\cite{zhang2019graph}.


Learning on multigraphs and heterogeneous networks supporting static data has been considered in previous works such as~\cite{zhang2018scalable,qu2017attention,fu2020magnn,cen2019representation,yu2022multiplex, 10.1145/3292500.3330961}, where the focus was on learning embeddings. For multigraphs supporting signals, GSP tools have been used to perform convolutional learning on each graph that composes the multigraph separately, without considering inter-graph flows of information~\cite{9564196, ke2021joint, wang2021forecasting}.


In this paper, we present multigraph convolutional networks (MGNNs) to perform learning with signals supported on multigraphs, relying on convolutional multigraph signal processing (MSP). This approach is aimed to exploit diffusions of heterogeneous, inter-graph information leveraging the benefits of convolutional operations. To derive the MSP model, we exploit algebraic signal processing (ASP), stating the notions of signals and filtering. Then, we stack layers where information is processed by means of MSP, and pointwise non-linearities are used to map information between layers. We provide a procedure for the efficient computation of multigraph filters, and a low cost method to reduce the dimension of the data transferred between the layers of the MGNN. We also perform a set of numerical experiments where we evaluate the performance of MGNNs against other architectures used to learn from information on multigraphs.


From our numerical experiments, we observe that it is possible to perform learning on multigraphs, and this can be done with arbitrary selections of the shift operators describing the individual graphs that constitute the multigraph. Additionally, we find that MGNNs exhibit a superior performance to graph neural networks when considered for the learning of information on multichannel wireless communication systems.




\section{MULTIGRAPH SIGNAL PROCESSING (MSP)}
\label{sec:msp}


\begin{figure}
	\centering
	\includegraphics[scale=1]{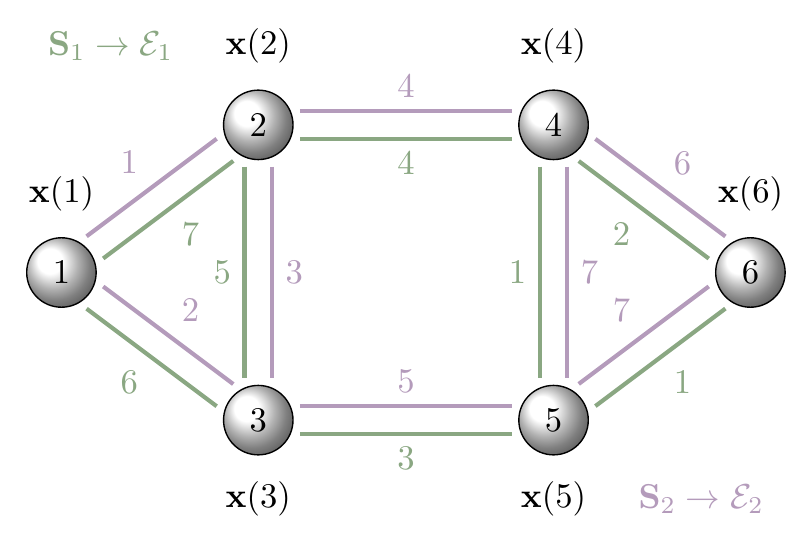} 
	\caption{Depiction of the multigraph $M = (\ccalV, \{ \ccalE_1 , \ccalE_2 \})$ and a signal $\bbx$ on $M$. The nodes in $\ccalV$ are connected by two types of edges, $\ccalE_1$ and $\ccalE_2$.}
	\label{fig:basicMultigraph}
\end{figure}


\begin{figure*}[t]
\centering
\includegraphics[width=\textwidth]{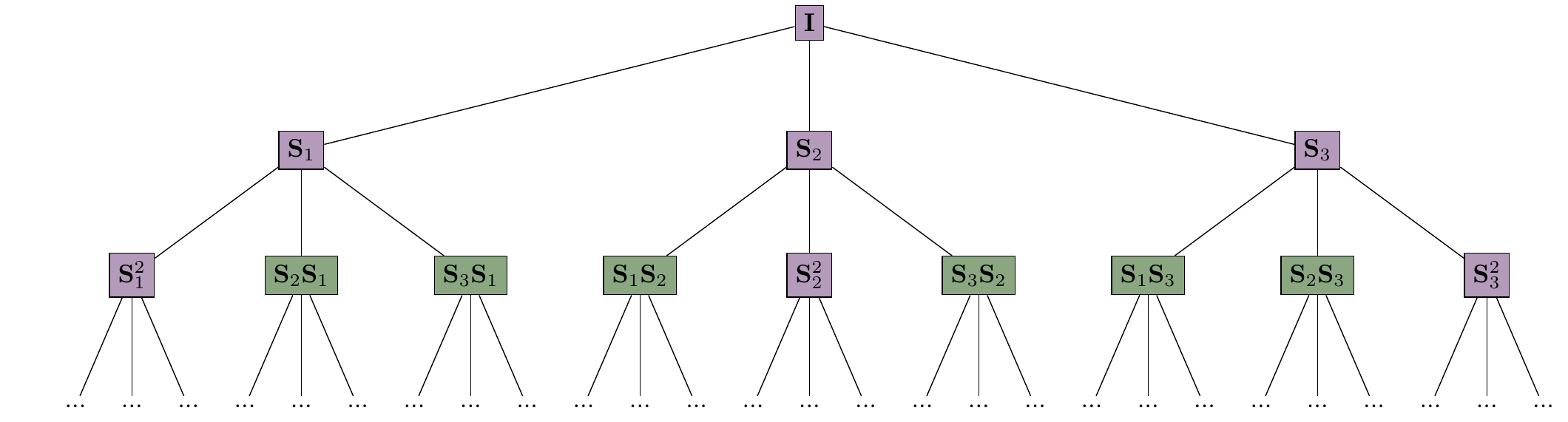}
\caption{Diffusion tree for multigraph filters with three shift operators $\mathbf{S}_1$, $\mathbf{S}_2$, and $\mathbf{S}_3$. Monomials in green are multivariate diffusions, which are not considered by traditional graph neural networks.}
\label{fig:diffTreeICASSP}
\end{figure*}

A multigraph $M$ is defined by the pair  $\left( \ccalV,  \{ \ccalE_i \}_{i=1}^{m} \right)$, where $\ccalV$ is a node set and $\{ \ccalE_i \}_{i=1}^{m} $ is a collection of edge sets, such that each  pair $(\ccalV, \ccalE_i)$ constitutes a graph with nodes in $\ccalV$ and edges in $\ccalE_i$~\cite{Butler2022ConvolutionalLO} -- see Fig.~\ref{fig:basicMultigraph}. The matrix representation of the multigraph $M$ will be given by the collection of matrices associated to the graphs $\{  (\ccalV, \ccalE_i) \}_{i=1}^{m}$. For instance, if $\bbW_i$ is the adjacency matrix of  $(\ccalV, \ccalE_i)$, the collection $\{  \bbW_i \}_{i=1}^{m}$ is the matrix representation of the multigraph $M$.

Like in the case of graphs, signals can be defined on multigraphs associating a scalar value to each node in $\ccalV$. Then, a multigraph signal can be identified with a vector in $\bbx\in\mbR^{N}$, where $N=\vert \ccalV\vert$. The $i$-th component of $\bbx$ is the scalar value defined on the $i$-th node in $\ccalV$ given a labeling of $\ccalV$ -- see Fig.~\ref{fig:basicMultigraph}. Notice that the set of signals on the multigraph $M$ constitute a vector space that we denote by $\ccalM$.

Graph signals are diffused on the graph by means of the action of the graph shift operator. In order to properly define diffusion on multigraphs, we introduce the definition of shift operators for multigraphs.


\begin{definition}[\hspace{-.013cm}\cite{Butler2022ConvolutionalLO}]\label{def_shift_operator_multigraph}
	
	Let $M=\left( \ccalV,  \{ \ccalE_i \}_{i=1}^{m} \right)$ be a multigraph, and let $\bbS_i$ be the shift operator of the graph $ (\ccalV, \ccalE_i) $. Then, each $\bbS_i$ is a shift operator of $M$.
	
\end{definition}


The diffusions of a multigraph signal $\bbx$ defined on the multigraph $M$ with shift operators $\{ \bbS_i \}_{i=1}^{m}$ have the form $\bbS_{i_1}^{k_1}\ldots\bbS_{i_r}^{k_r}\bbx$, where $i_\ell \in \{ 1,\ldots, m \}$ and $k_\ell \in\mbN$. Under this form, information can diffuse not only within each edge set of the multigraph but also between edge sets. It is important to remark that the selection of the shift operators $\bbS_i$ is inter-graph independent, such that the choice of $\bbS_i$ is independent from the choice of $\bbS_j$ for $i\neq j$. With these concepts at hand, we formally define multigraph filters as follows.


\begin{definition}[\hspace{-.013cm}\cite{Butler2022ConvolutionalLO}]
	
	Let $M= \left( \ccalV, \{ \ccalE_i \}_{i=1}^{m}  \right)$ be a multigraph with shift operators $\{ \bbS_i \}_{i=1}^{m}$. Then, a multigraph filter on $M$ is a polynomial function $p: \text{End}(\ccalM)^m \rightarrow \text{End}(\ccalM)$ denoted by $p(\bbS_1,\ldots,\bbS_m)$ whose independent variables are the shift operators $\bbS_i$. Here, $\text{End}(\ccalM)$ is the space of endomorphisms of $\ccalM$ onto itself.
	
\end{definition}


Then, if $\bbx$ is a signal on the multigraph with shift operators $\{ \bbS_i \}_{i=1}^{m} $, we say that $\bby = p(\bbS_1,\ldots,\bbS_m)\bbx$ is a filtered version of $\bbx$ by means of the filter $p(\bbS_1,\ldots,\bbS_m)$. Notice that the operators $\bbS_i$ do not necessarily commute.

In Fig.~\ref{fig:diffTreeICASSP}, we portray the structure behind diffusions associated to multigraph filters with three shift operators $\bbS_i$ for $i=1,2,3$. These diffusions represent the basis of filters for three shift operators, such that any multigraph filter can be written as a linear combination of these diffusions.  Graph filters, and thus graph neural networks, do not consider multivariate diffusions such as $\{ \bbS_i \bbS_j \}_{i,j=1, i\neq j}^{3}$ in their bases.


\subsection{Convolutional Attributes of Multigraph Filters}

Now, we will show why the notion of multigraph filtering is formally convolutional in the sense of~\cite{algSP0}. We recall that any convolutional signal model (ASM) in the light of algebraic signal processing (ASP) is defined by the triplet
$
(\ccalA , \ccalM, \rho)
,
$
where $\ccalA$ is an algebra, $\ccalM$ is a vector space, and $\rho$ is a homomorphism from $\ccalA$ into the space of endomorphisms of $\ccalM$~\cite{algSP2,algSP4,algSP5, parada_quiversp,alej2020graphon,C2,algnn_nc_j}. Notice that $\ccalA$ is a vector space where there is a notion of product that is closed. 

To show that multigraph filtering on $ M = \left( \ccalV , \{ \ccalE_i \}_{i=1}^{m}\right) $ is indeed convolutional, we choose $\ccalM = \mbR^{N}$ with $N = \vert\ccalV\vert$, while the algebra $\ccalA$ is a polynomial algebra with generators, $\{ g_i \}_{i=1}^{m}$. Then, the elements of $\ccalA$ are multivariate polynomials with independent variables given by $\{ g_i \}_{i=1}^{m}$, and where the  $g_i$-s are not necessarily commutative. It is important to note that the elements of $\ccalA$ are \textit{abstract} polynomials, where the independent variables are not taking values in any specific space or field,  as such, $\ccalA$ is a \textit{free} algebra~\cite{algnn_nc_j}. Finally, we choose a homomorphism $\rho: \ccalA \to \text{End}(\ccalM)$ given by $\rho(g_i) = \bbS_i$ for all $i=1,\ldots, m$. Notice that the translation of the filters in $\ccalA$ into operators in $\text{End}(\ccalM)$ is not a trivial substitution of variables, but instead is an implementation of the convolutional rules embedded in $\ccalA$. For instance, when $g_i g_j \neq g_j g_i$, it is not possible to know how far is $g_i g_j$ from $ g_j g_i $, but it is quantifiable for $\bbS_i \bbS_j$ and $ \bbS_j \bbS_i $.

Spectral decompositions of multigraph filters descend from the notion of irreducible subrepresentation. This translates into joint block diagonal decompositions of the filters and shift operators $\bbS_i$ defined on the multigraph. If the $\{  \bbS_i \}_{i=1}^{m}$ do not commute, some frequencies may be associated to matrices instead of scalars. For more details about the spectral representation of multigraph filters see~\cite{Butler2022ConvolutionalLO}.


\subsection{Efficient Computation of Multigraph Filters}


Multigraph filters are polynomial functions of the shift operators $\bbS_i$. The enumeration of all combinations of products between the $\bbS_i$ scales quickly with the number of shift operators and diffusion order -- see Fig.~\ref{fig:diffTreeICASSP}. To reduce the computational burden,  we propose a pruning method that reduces the number of  monomials in the filters by grouping those terms that depend on operator compositions that are close to being commutative. 

More precisely, if the commutator $\Vert \bbS_i \bbS_j - \bbS_j \bbS_i \Vert_2 \leq \epsilon\ll 1$, we retain monomials containing the term $\bbS_i \bbS_j$ while removing monomials containing $\bbS_j \bbS_i$.  Under the assumption that the individual shift operators are normalized, we have the following bound on pruned monomials.

\begin{lemma}[\hspace{-.013cm}\cite{Butler2022ConvolutionalLO}]\label{lemma_pruning_bound}

Let $M$ be a multigraph with $m$ shift operators $\{ \bbS_i \}_{i=1}^{m}$, where $\Vert \bbS_i \Vert_2 \leq 1$ for all $i$. Let $[\bbS_i , \bbS_j] = \bbS_i \bbS_j - \bbS_j \bbS_i $ with $\Vert  [\bbS_i , \bbS_j] \Vert_2 \leq \epsilon\ll 1$ for some $i,j$. Then, it follows that
$
\left\Vert 
      \bbS^{k_1}_{i_1}\cdots\bbS^{k_m}_{i_m}
      [\bbS_i , \bbS_j]
      \bbS_{j_1}^{\ell_1}\cdots\bbS_{j_m}^{\ell_m}
\right\Vert_2
\leq 
\epsilon
$
for all $i_r , j_r \in \{ 1,\ldots, m \}$ and $k_r, \ell_r \in\mbN$. 
\end{lemma}


When two shift operators $\bbS_i$ and $\bbS_j$ are nearly commutative by an error of $\epsilon$, the monomials of the filter containing $\bbS_i \bbS_j$ and monomials containing $\bbS_j \bbS_i$ differ up to a factor of $\epsilon$. We remark that since the proposed approach relies on having a value of $\epsilon\ll 1$, then the pruning method is as accurate as the value of $\epsilon$ allows it. The complete procedure for computing the pruned, depth-limited filters is provided in Algorithm 2 of  \cite{Butler2022ConvolutionalLO}.

\section{MULTIGRAPH CONVOLUTIONAL NEURAL NETWORKS
(MGNN)}
\label{sec:mgnn}

\begin{figure}[t]
\centering
\input{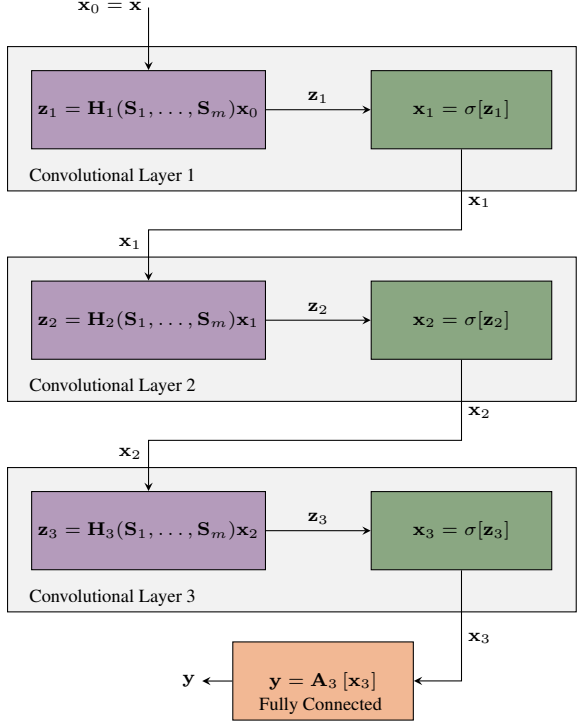}
\caption{Block diagram of a multigraph convolutional neural network with three convolutional layers and a fully connected layer.}
\label{fig:blockDiag}
\end{figure}

In this section, we introduce a multigraph convolutional neural network architecture that generalizes and extends the convolutional processing of information supported on a multigraph. We also present techniques for dimensionality reduction through selection sampling and pooling and conclude by showing that the architecture is equivariant to node relabelings. 

\subsection{Multigraph Perceptrons}
Multigraph convolutions enable a parameterization to map from an input signal $\bbx$ to  a target representation $\bby$. Given a training set $\mathcal{T}$ consisting of samples $(\bbx, \bby)$, we can learn filters coefficients corresponding to each monomial. For negligible computational increase, we can apply a pointwise nonlinearity $\sigma(\cdot)$ to the output of the multigraph convolution. We refer to this as the \emph{multigraph perceptron}, expressed as $ \sigma(\mathbf{H}(\mathbf{S}_1, \dots, \mathbf{S}_m)\mathbf{x})$, where $\mathbf{H}(\mathbf{S}_1, \dots, \mathbf{S}_m)$ is the multigraph filter.

To measure the success of a multigraph perceptron, consider a loss function $J(\cdot)$. Leveraging this loss function, optimal filters $\mathbf{H}^*$ are found through
\begin{equation}
    \label{eq:ERM_MG}
    \mathbf{H}^* = \argmin\limits_{\mathbf{H}}\hspace{0.25em}|\mathcal{T}|^{-1}\sum_{(\bbx,\bby)\in \mathcal{T}} J(\mathbf{y}, \sigma(\mathbf{H}(\mathbf{S}_1, \dots, \mathbf{S}_m)\bbx)).
\end{equation}

The output of a multigraph perceptron is another multigraph signal. Thus, we can compose $L$ layers of perceptrons where for each layer $\ell$, we process by
\begin{equation}
    \mathbf{x}_\ell = \sigma(\mathbf{H}_\ell(\mathbf{S}_1, \dots, \mathbf{S}_m)\mathbf{x}_{\ell - 1}).
\end{equation}


\begin{figure*}[t]
\hspace{0.15cm}
\begin{minipage}[b]{0.5\linewidth}
  \centering
  \centerline{\includegraphics[width=\linewidth]{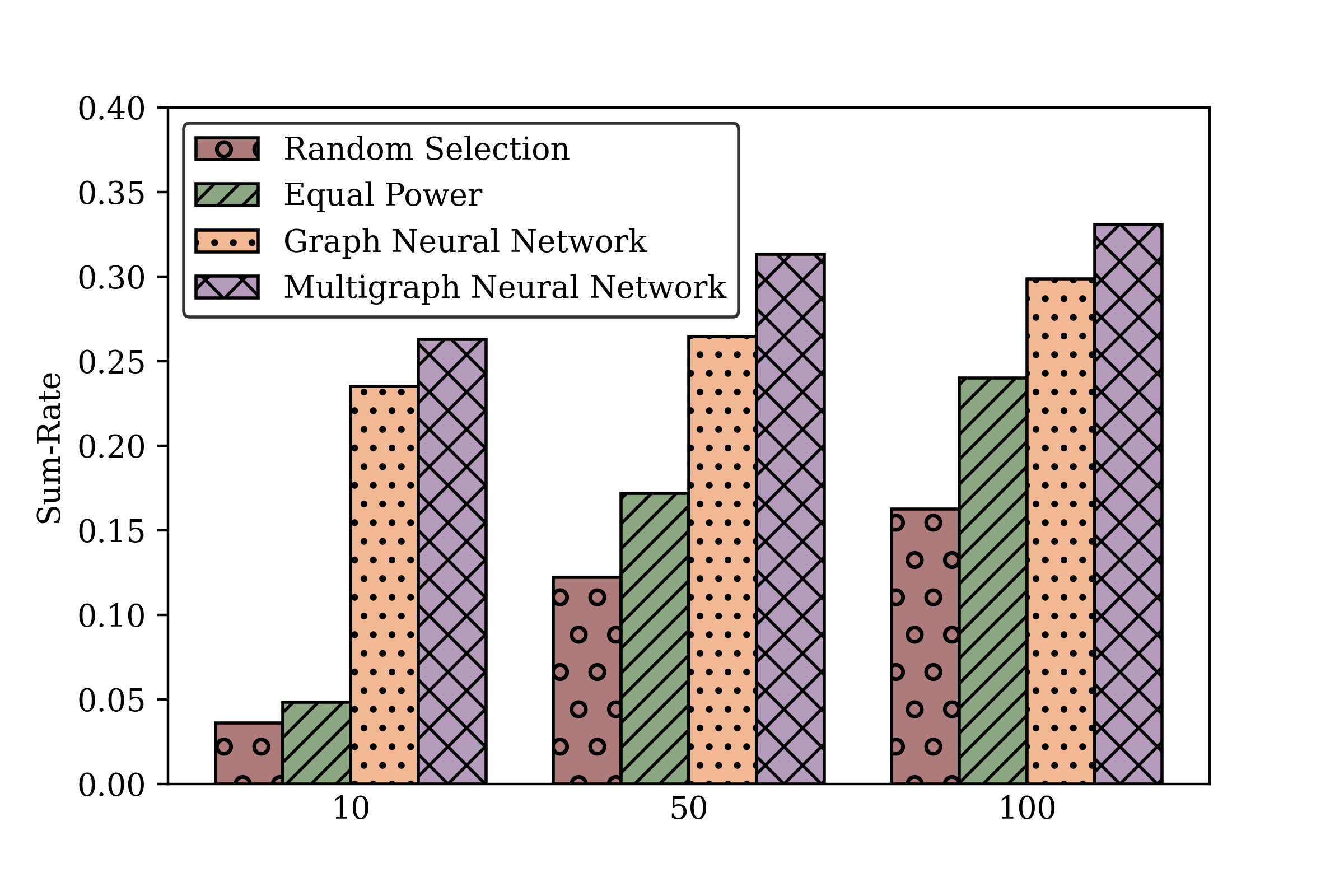}}
    \label{fig:powBudget}
  \centerline{(a) Power Budget}\medskip
\end{minipage}
\begin{minipage}[b]{0.5\linewidth}
  \centering
  \centerline{\includegraphics[width=\linewidth]{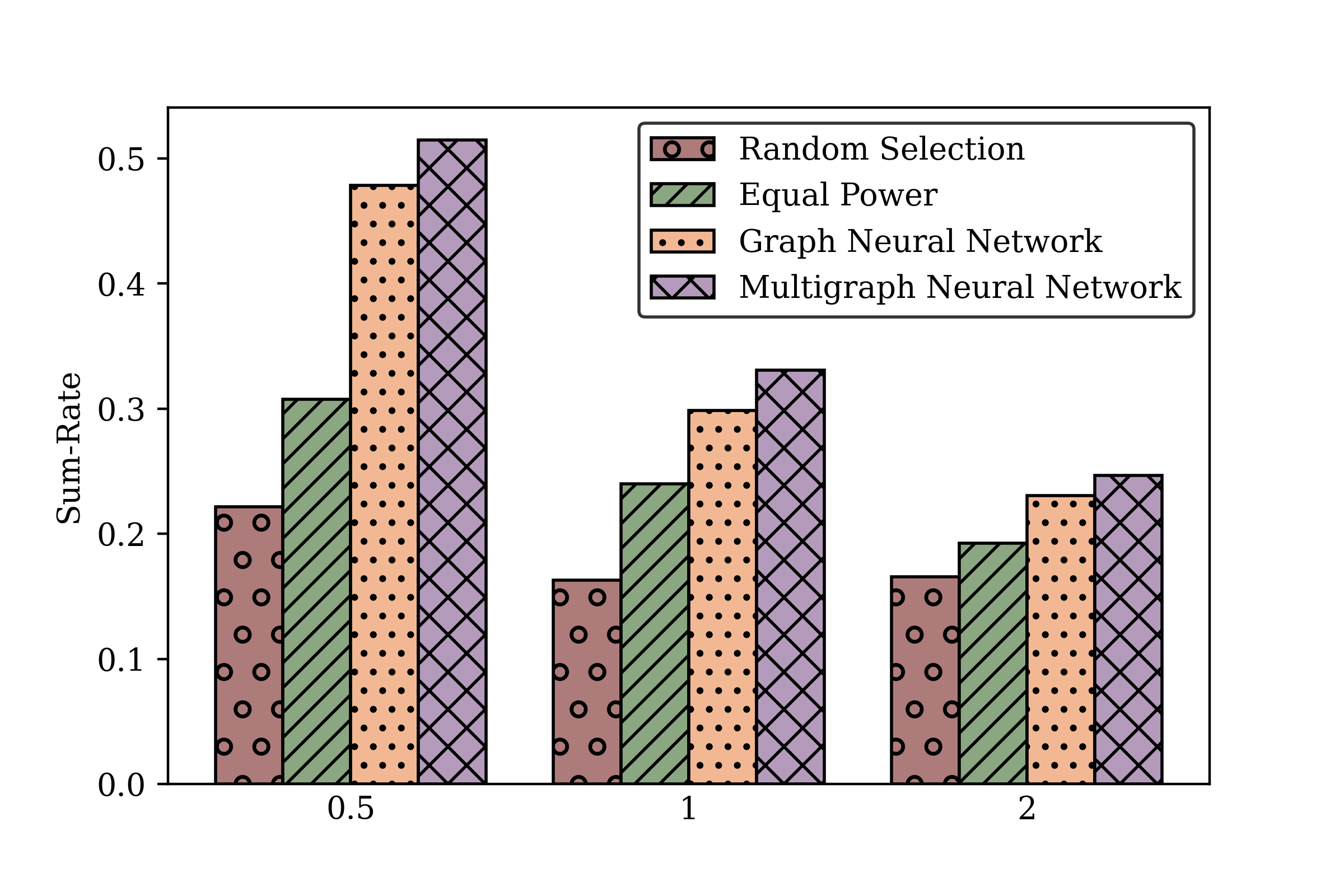}}

  \centerline{(b) Noise}\medskip
\end{minipage}
\caption{In (a), the power budget is altered from 10mW to 100mW. For (b), the
channel noise is varied from $0.5\cdot 10^{-3}$mW to $2\cdot 10^{-3}$mW.}
\label{fig:res}
\end{figure*}

The last layer can be appended by multiple fully connected layers $\mathbf{A}_L$ to reshape $\mathbf{x}_L$ to the target representation $\mathbf{y}$. The stacking of multigraph perceptrons combined with fully connected layers defines the \emph{multigraph convolutional neural network (MGNN)}. A complete block diagram of this architecture is illustrated in Fig. \ref{fig:blockDiag}.

\subsection{Selection Sampling and Pooling}
Pooling is useful for reducing dimensionality while preserving feature information. We extend pooling to the multigraph setting by generalizing a pooling technique for graph neural networks that has shown much empirical success. 

In \cite{gama2018convolutional}, the authors propose a procedure that at each layer, nodes are selected that best preserve the graph's original underlying properties and structure. For multigraphs, we can use centrality measures to identify such nodes, such as betweenness \cite{sole2014centrality} or PageRank \cite{halu2013multiplex}. 

Let $\mathcal{V}_\ell$ be the set of nodes selected at each layer $\ell$, with $|\mathcal{V}_\ell| = N_\ell$. If a node is selected at layer $\ell$, it must also be selected for all layers up to $\ell$. If we recorded the label of nodes such that those selected at layer $\ell$ always come before those that are not selected, we can extract the signals of selected nodes with a $N_{\ell} \times N_{\ell-1}$ wide binary sampling matrix $\mathbf{D}_{\ell}$ containing ones across the main diagonal and zeros elsewhere. Then, the shift operator for the layer $\ell$ and graph $g$ is given by
\begin{equation}
\mathbf{S}_{\ell,g} =\mathbf{D}_{\ell} \mathbf{S}_{\ell-1,g} \mathbf{D}^\intercal_{\ell},
\end{equation}
as well as sampled signals given by
\begin{equation}
\tilde{\mathbf{x}}_{\ell-1} = \mathbf{D}^\intercal_{\ell}\mathbf{x}_{\ell-1},
\end{equation}
where $\mathbf{S}_{0,g} = \mathbf{S}_{g}$ and $\mathbf{x}_{0} = \mathbf{x}$.
We this sampling procedure, we can express the output of the multigraph perceptron as
\begin{equation}
 \begin{split}
    \mathbf{x}_\ell &= \sigma(\mathbf{H}_\ell(\mathbf{S}_{\ell,1}, \dots, \mathbf{S}_{\ell,m})\tilde{\mathbf{x}}_{\ell - 1}).\\
 \end{split}
\end{equation}

To preserve feature information of non-selected nodes, we pool community feature information onto selected nodes. Let $\mathbf{n}_{\ell,i,g}$ be the set of nodes reachable from node $i$ on graph $g$ within $\alpha_{\ell}$ steps for each layer ${\ell}$. At the first layer, this is the set of nodes $j$ with non-zero entries $\left[\mathbf{S}_g^k\right]_{ij} \neq 0$ for all integer powers $k$ up to $\alpha_{\ell}$. For successive layers, we compute these sets through use of a $N_{\ell} \times N$ wide binary sampling matrix $\mathbf{E}_{\ell}$ also containing ones across the main diagonal and zeros elsewhere
\begin{equation}
    \mathbf{n}_{\ell,i,g} = \left\{j : \left[\mathbf{E}_{\ell}\mathbf{S}_{g}^k\mathbf{E}^\intercal_{\ell - 1}\right]_{ij} \neq 0 \text{ for } k = 0, \dots, \alpha_{\ell} \right\}.
\end{equation}

The multigraph neighborhood $\mathbf{n}_{\ell,i}$ at layer $\ell$ is simply defined as the union of each of the graph neighborhoods
\begin{equation}
    \mathbf{n}_{\ell,i} = \bigcup\limits_{g=1}^{m} \mathbf{n}_{\ell,i,g}.
\end{equation}

To preserve feature information within $i$'s locality, traditional aggregation operations such as the  mean, median, or maximum over each feature can be applied to $\mathbf{n}_{\ell,i}$. The complete multigraph convolutional neural network (MGNN) algorithm with selection and pooling is provided in Alg. 1 of \cite{Butler2022ConvolutionalLO}.

\subsection{Label Permutation Equivariance}

 Multigraph neural networks are equivariant to permutations of node labelings, such that processing occurs independently of labeling. Consider an arbitrary permutation matrix $\mathbf{P}$ such that $\mathbf{P}^\intercal\mathbf{P}=\mathbf{P}\mathbf{P}^\intercal=\mathbf{I}$. Applying a reordering of nodes will generate a permuted signal $\hat{\mathbf{x}}_{\ell - 1} = \mathbf{P}^\intercal\mathbf{x}_{\ell - 1}$ and permuted shift operators $\hat{\mathbf{S}}_i = \mathbf{P}^\intercal\mathbf{S}_i\mathbf{P}$. We are now able to state the following lemma.


\begin{lemma}[\hspace{-.02cm}\cite{Butler2022ConvolutionalLO}] Given a consistent relabeling of signals and shift operators by permutation matrix $\mathbf{P}$, the outputs of each layer $\mathbf{x}_\ell \coloneqq \sigma(\mathbf{H}_\ell(\mathbf{S}_1, \dots, \mathbf{S}_m)\mathbf{x}_{\ell - 1})$ and $\hat{\mathbf{x}}_\ell \coloneqq \sigma(\mathbf{H}_\ell(\hat{\mathbf{S}}_1, \dots, \hat{\mathbf{S}}_m)\hat{\mathbf{x}}_{\ell - 1})$ satisfy 
$	
\hat{\mathbf{x}}_{\ell} := \mathbf{P}^\intercal \mathbf{x}_{\ell}.
$	
%
%
%
%
\end{lemma}


For proof, see \cite{Butler2022ConvolutionalLO} Lemma 2.

\section{Numerical EXPERIMENTS}
\label{sec:exp}

In this section, we compare the proposed MGNN architecture against heuristics and traditional graph neural networks on an optimal resource allocation task over a multi-band wireless communication network \cite{Butler2022ConvolutionalLO}. Consider a wireless system consisting of $40$ transmitters each assigned to one of $10$ receivers. Receivers are placed uniformly at random with positions $\mathbf{r}_i \in [-40,40]^2$ and associated transmitters placed randomly nearby at $\mathbf{t}_i \in [\mathbf{r}_i - 10, \mathbf{r}_i + 10]$. The channel between each transmitter and receiver is composed of a constant path loss gain and a time varying fast fading factor. 

Transmitters may communicate over the 2.4GHz and 5GHz frequency bands and must decide how much power they devote to each. All transmitters are subject to a shared maximum average power budget and seek to maximize the expected sum-rate (i.e. the expected sum of all receiver capacities over both frequency bands). The complete experimental setup is provided in Section IV.A of \cite{Butler2022ConvolutionalLO}.

The two heuristics we consider are random selection (randomly selecting half of the transmitters to communicate with equal power over each frequency band) and equal power (all transmitters assign equal power to each transmitter and frequency band).
As an extension of \cite{9072356} to the multi-band setting, we evaluate a multi-channel graph neural network with two shift operators corresponding to the channel strength over each frequency band. These same inputs are fed into a multigraph neural network. Both architectures contain two convolutional layers, each with a maximum diffusive order of 3, 2 filters, and a ReLU activation function. We employ the primal-dual learning method detailed in \cite{9072356} to solve the constrained learning problem for the optimal filters.

\vspace{-0.25cm}
To study the performance of the four algorithms, we vary the system's power budget and level of noise \cite{Butler2022ConvolutionalLO}. As shown in Fig. \ref{fig:res}a, the MGNN achieves the best sum-rate for systems with power budgets of 10mW, 50mW, and 100mW. Fig.   \ref{fig:res}b showcases the advantage of the multigraph neural network in different SINR regimes. Even after altering the white Gaussian noise of the channel to range between $\{0.5,1,2\}\cdot 10^{-3}$mW, the MGNN offers a strong improvement upon the heuristics and the graph neural network. These results validate the claim that by considering a broader class of dynamics,  multigraph neural networks can outperform graph neural networks when applied to systems that have a natural multigraph modeling.

\section{Conclusions}
\label{sec:conc}

We introduced convolutional signal processing on multigraphs (MSP) exploiting the theory of algebraic signal processing (ASP). Using MSP, we proposed multigraph convolutional neural networks (MGNNs) as stacked layered structures where information is processed in each layer by means of a MSP model. MGNNs allow the learning of filters that can capture mixed heterogeneous diffusions between the graphs that constitute a multigraph, while still preserving the convolutional attributes in the filtering operations. To provide an efficient computation of the filter coefficients, we presented an operator pruning method that reduces the size of the set of allowable filters. To reduce the dimension of growing data between layers, we proposed a pooling approach aimed at preserving structural properties of the signals. The numerical experiments performed on a multi-channel wireless communication system show that MGNNs are superior to GNNs for learning on multigraphs. This is rooted on the fact that the convolutions in MGNNs capture unique inter-graph flows of information that are not captured by GNNs.


\pagebreak
\bibliographystyle{IEEEbib}
\bibliography{bibliography}

\end{document}